\documentclass[sigplan,screen]{acmart}

\usepackage{colortbl}
\usepackage{tikz}
\usepackage{xspace}
\usepackage{siunitx}
\usepackage{amsmath}
\usepackage{subfig}
\usepackage{subfloat}
\usepackage{soul}
\usepackage{textcomp}

\AtBeginDocument{%
  }

\copyrightyear{2025}
\acmYear{2025}
\setcopyright{cc}
\setcctype{by}
\acmConference[ASPLOS '25]{Proceedings of the 29th ACM International Conference on Architectural Support for Programming Languages and Operating Systems, Volume 1}{March 30-April 3, 2025}{Rotterdam, Netherlands}
\acmBooktitle{Proceedings of the 29th ACM International Conference on Architectural Support for Programming Languages and Operating Systems, Volume 1 (ASPLOS '25), March 30-April 3, 2025, Rotterdam, Netherlands}\acmDOI{10.1145/3669940.3707227}
\acmISBN{979-8-4007-0698-1/25/03}
\newcommand{\revise}[1]{#1}

\newcommand*\circled[2]{\protect\tikz[baseline=(char.base)]{
            \protect\node[shape=circle,fill=black,inner sep=1pt] (char) {\textcolor{#1}{{\footnotesize #2}}};}}

\ifx\figurename\undefined \def\figurename{Figure}\fi
\renewcommand{\figurename}{Fig.}
\renewcommand{\paragraph}[1]{\textbf{#1} }

\newcommand{\Sect}[1]{Sec.~\ref{#1}}
\newcommand{\Fig}[1]{Fig.~\ref{#1}}
\newcommand{\Tbl}[1]{Tbl.~\ref{#1}}
\newcommand{\Eqn}[1]{Eqn.~\ref{#1}}

\newcommand{\proj}{\textsc{MetaSapiens}\xspace}

\newcommand{\mode}[1]{\underline{\textsc{#1}}\xspace}

\newcommand{\RNum}[1]{\uppercase\expandafter{\romannumeral #1\relax}}

\newcommand{\cL}{\mathcal{L}}
\newcommand{\cM}{\mathcal{M}}

\graphicspath{{imgs/}}

\begin{document}

\title[\proj: Real-Time Neural Rendering with Efficient Pruning and Foveated Rendering]{\proj: Real-Time Neural Rendering with Efficiency-Aware Pruning and Accelerated Foveated Rendering}

\author{Weikai Lin}
\authornote{Both authors contributed equally to this research.}
\orcid{0000-0003-3537-4857}
\affiliation{%
  \institution{University of Rochester}
  \city{Rochester}
  \state{NY}
  \country{USA}
}
\email{wlin33@ur.rochester.edu}

\author{Yu Feng}
\authornotemark[1]
\orcid{0000-0002-2192-5737}
\affiliation{%
  \institution{Shanghai Jiao Tong University}
  \city{Shanghai}
  \country{China}
}
\email{y-feng@sjtu.edu.cn}

\author{Yuhao Zhu}
\orcid{0000-0002-2802-0578}
\affiliation{%
  \institution{University of Rochester}
  \city{Rochester}
  \state{NY}
  \country{USA}
}
\email{yzhu@rochester.edu}

\begin{abstract}
Point-Based Neural Rendering (PBNR) \revise{is emerging as}
a promising class of rendering techniques, which are permeating all aspects of society, driven by a growing demand for real-time, photorealistic rendering in AR/VR and digital twins. 
Achieving real-time PBNR on mobile devices is challenging.

This paper proposes \proj, a PBNR system that for the first time delivers real-time neural rendering on mobile devices while maintaining human visual quality.
\proj combines three techniques.
First, we present an efficiency-aware pruning technique to optimize rendering speed. 
Second, we introduce a Foveated Rendering (FR) method for PBNR, leveraging humans' low visual acuity in peripheral regions to relax rendering quality and improve rendering speed.
Finally, we propose an accelerator design for FR, addressing the load imbalance issue in (FR-based) PBNR.
Our evaluation shows that our system achieves an order of magnitude speedup over existing PBNR models without sacrificing subjective visual quality, as confirmed by a user study. \revise{The code and demo are available at: \href{https://horizon-lab.org/metasapiens/}{https://horizon-lab.org/metasapiens/}.}
\end{abstract}



\maketitle

\section{Introduction}
\label{sec:intro}

Rendering technologies are infiltrating every facet of society.
For instance, rendering is critical to enabling digital twins~\cite{juarez2021digital} in emerging areas such as smart cities, digital healthcare, and telepresence.
Reinvigorated interests in Augmented/Virtual Reality (AR/VR) further heighten the demand for real-time, photorealistic rendering.

Point-Based Neural Rendering (PBNR), i.e., the Gaussian Splatting-family algorithms~\cite{Kerbl2023GaussianSplatting, fan2023lightgaussian, lee2024compact, fang2024mini, niemeyer2024radsplat, girish2023eagles}, \revise{is emerging as} 
a new class of rendering solutions, which revitalizes the classic point-based rendering techniques~\cite{gross2011point, levoy1985use, pfister2000surfels, zwicker2001surface} using modern neural rendering methods~\cite{mildenhall2021nerf}.
PBNR, like previous neural rendering algorithms such as Neural Radiance Fields (NeRF), offers photorealistic rendering by learning \revise{scene radiance}
from data, but is significantly faster by replacing the compute-intensive Multilayer Perceptrons (MLPs) in NeRF with lightweight point-based rasterization.

Nevertheless, PBNR is still far from real-time on mobile devices, rendering generally below 10 Frames-Per-Second (FPS) on the mobile Volta GPU~\cite{xaviersoc}.
This paper introduces \proj, which, for the first time, delivers real-time PBNR on mobile devices while maintaining human visual quality.
\proj combines three key ingredients.

\paragraph{Efficiency-Aware Pruning.}
Much of the recent efforts on optimizing PBNR models focus on pruning~\cite{fan2023lightgaussian, lee2024compact, fang2024mini, niemeyer2024radsplat, girish2023eagles}, which, while reducing the model size, does not bring significant speedups.
This is because existing pruning methods are single-minded in reducing the sheer number of points while being agnostic to the actual computational cost.
We find that different points in a PBNR model contribute differently to the overall computation.
Instead, we propose an efficiency-aware pruning method that directly optimizes for the rendering/inference speed (\Sect{sec:prune}).

\paragraph{Foveated PBNR.}
\proj also exploits characteristics of human vision to improve performance (\Sect{sec:fr}).
Human vision acuity is poor in the visual periphery~\cite{wandell1995foundations}, an opportunity that has long been exploited: one can speed up rendering by gradually reducing the rendering quality as the pixel eccentricity increases (i.e., as pixels are positioned more at the visual periphery) with impunity, a technique known as Foveated Rendering (FR)~\cite{patney2016towards, guenter2012foveated}.

We introduce the first FR method for PBNR.
We gradually reduce the number of points used for rendering as the pixel eccentricity increases.
The key is a data representation, where points at higher eccentricies are purposely designed to be a strict subset of the points at lower eccentricies.
That way, (most of the) parameters and computation are shared when rendering different eccentricity regions, improving performance and reducing storage requirements.

Equally important to improving performance is to maintain a high visual quality.
To that end, we introduce a new training method, which guides both pruning and peripheral quality relaxation (in FR) by explicitly modeling human visual perception at different eccentricies.
As a result, the subjective visual quality is consistent across the visual field and is aligned with the dense, non-FR model.

\paragraph{Architectural Support.}
While the two techniques above readily provide about an order of magnitude speedup on GPU,
we co-design an accelerator architecture to further improve the performance (\Sect{sec:hw}).
Aside from providing hardware support for FR,
the hardware addresses a key performance bottleneck in PBNR exacerbated by FR: low hardware utilization due to workload imbalance across tiles in a frame.
Our hardware mitigates this issue via 1) a dynamic tile merging scheme that balances workloads across tiles and 2) incrementally pipelining adjacent stages through line-buffering.

\paragraph{Result.}
We evaluate our method using both subjective human studies and objective measurements of performance and quality.
Across 12 participants, the subjective rendering quality of our method is statistically no-worse than that of Mini-Splatting-D ~\cite{fang2024mini}, a state-of-the-art PBNR method in rendering quality.
Compared to five state-of-the-art PBNR methods, we out-perform all of them in both objective rendering quality (by up to 0.4 dB in PSNR) and rendering speed (by up to 7.4$\times$ on a mobile Volta GPU and 20.9$\times$ with our hardware support).

The contributions of this paper are as follows:
\begin{itemize}
\item We propose an efficiency-aware pruning method for PBNR that directly targets computational efficiency rather than merely reducing point counts.
\item We propose the first FR method tailored for PBNR; the method is centered around a new data/point representation, which improves rendering performance with little storage overhead.
\item We introduce a training framework that incorporates both pruning and FR while maintaining subjective visual quality; the key is to explicitly model HVS and use the model to guide training.
\item We co-design an accelerator architecture, which addresses the load imbalance issue in PBNR and futher improves performance.
\end{itemize}

\section{Background}
\label{sec:bck}

We first introduce the necessary background in PBNR (\Sect{sec:bck:pbnr}), followed by the main characteristics of the Human Visual System (HVS) and how they are used by Foveated Rendering to improve rendering speed (\Sect{sec:bck:hvs}).

\begin{figure}[t]
    \centering
    \includegraphics[width=\columnwidth]{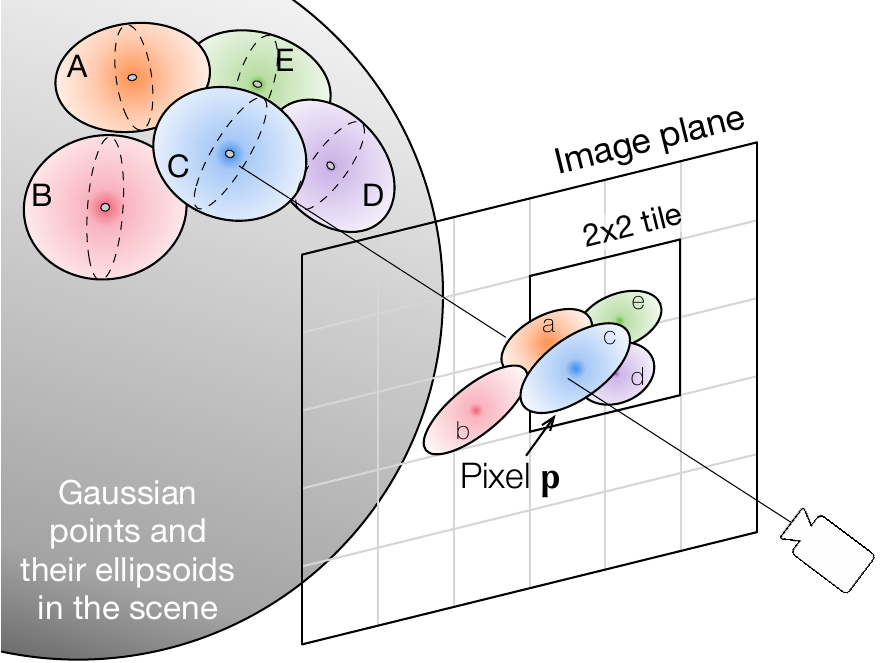}
\caption{
Illustration of PBNR, which parameterizes the scene with a set of points, each associated with a 3D Gaussian distribution that gives rise to an ellipsoid.
The ellipsoids are projected to ellipses on the image plane, where the ellipses are sorted (per tile, e.g., $2\times 2$ pixels).
The color of a pixel is calculated by integrating the contribution of each intersecting ellipse (e.g., \texttt{a}, \texttt{c}, \texttt{d}, \texttt{e} for \textbf{p}).
}
    \label{fig:pbnr}
\end{figure}

\subsection{Point-Based Neural Rendering}
\label{sec:bck:pbnr}

PBNR is a class of neural rendering techniques, exemplified by the 3D Gaussian Splatting (3DGS) algorithm~\cite{Kerbl2023GaussianSplatting} and its descendants~\cite{fan2023lightgaussian, lee2024compact, fang2024mini, niemeyer2024radsplat, girish2023eagles}.
Compared to previous neural rendering techniques, a.k.a., the NeRF-family algorithms~\cite{mildenhall2021nerf, barron2022mip, muller2022instant, chen2022tensorf, sun2022direct}, PBNR is fundamentally more efficient (e.g., usually over 1,000 times faster), because it parameterizes the scene with discrete points (rather than voxels) to avoid redundant computations and renders via a lightweight rasterization-based process called splatting~\cite{rusinkiewicz2000qsplat, zwicker2001surface, ren2002object} rather than the heavy MLP inference.

\paragraph{General PBNR Pipeline.}
We use 3DGS as a running example to explain the general pipeline of PBNR, which all PBNR algorithms follow.
\Fig{fig:pbnr} illustrates the general idea.
Rendering starts with an offline-trained model, which contains a set of discrete points (\texttt{A}--\texttt{E}) that represent the scene.
Each point is associated with an ellipsoid, whose three-dimensional scales are determined by the $\sigma$s of a 3D Gaussian distributions (hence the name Gaussian point).
Each ellipsoid has a set of \textit{trainable} parameters, including the scales, position, orientation, opacity, color distribution (which is parameterized through Spherical Harmonics; SH).

Given the trained points/ellipsoids,
the online rendering follows three steps: \textit{Projection}, \textit{Sorting}, and \textit{Rasterization}.

\underline{Projection.}~
Each ellipsoid is first projected/splatted to an ellipse on the image plane\footnote{We use ``points'', ``ellipses'', and ``ellipsoids'' interchangeably: there is a one-to-one mapping between them.}.
In the example of \Fig{fig:pbnr}, the ellipsoids \texttt{A}--\texttt{E} in the scene are splatted to ellipses \texttt{a}--\texttt{e} on the image plane.
The goal is to identify, for each pixel tile (e.g., $2\times 2$), which ellipses intersect with the tile and thus contribute to the pixel colors in the tile.

\underline{Sorting.}~ 
For each tile, we sort all the intersecting ellipses based on their depths to the image plane;
that way, closer ellipses can be \revise{integrated first}  
when calculating pixel colors.
For instance in \Fig{fig:pbnr}, ellipse \texttt{c} would be the closest.

\underline{Rasterization.}~
Finally, we calculate the intersections of all the ellipses in a tile with each pixel.
The color of a pixel $\textbf{p}$ is then computed using the classic volume rendering method~\cite{max1995volume_rendering, kaufman1993volume, levoy1988display}, which integrates the contribution of all the intersecting ellipses from near to far:
\begin{subequations}
\label{eqn:alphablending}
\begin{align}
\textbf{p} = \sum_{i=0}^{N-1} T_i \alpha_i c_i, ~~~ T_i = \prod_{j=0}^{i-1} (1 - \alpha_j ) \label{eq:p}\\
\alpha_i = f(\text{opacity}_i, ..., \text{pose}) \label{eq:f}\\
c_i = g(\text{SH}_0, ..., \text{SH}_n) \label{eq:g}
\end{align}
\end{subequations}

\noindent where $N$ is the number of ellipses intersecting $\textbf{p}$ (i.e., \texttt{a}, \texttt{c}, \texttt{d}, \texttt{e} in \Fig{fig:pbnr}), $\alpha_i$ is a function $f$ of various trainable parameters of the $i^{th}$ intersecting ellipse (e.g., opacity) and the camera pose, and $c_i$ is the color of the $i^{th}$ ellipse at the position of intersection, which is calculated using the Spherical Harmonics (SH) function $g$ with trainable coefficients $\text{SH}_{n}$.
We refer readers to Kerbl et al.~\cite{Kerbl2023GaussianSplatting} for the details of $f$ and $g$.

\subsection{Human Visual System and Foveated Rendering}
\label{sec:bck:hvs}

\begin{figure}[t]
    \centering
    \includegraphics[width=\columnwidth]{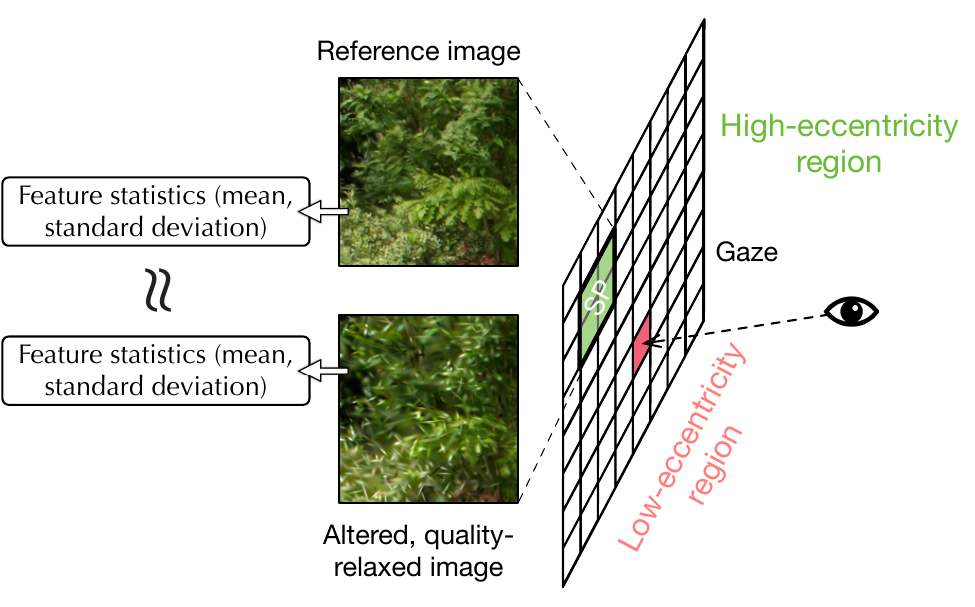}
\caption{
Pixels under the user's gaze have low eccentricities, where the human visual quality is the highest;
the peripheral pixels have high eccentricities where human visual acuity is low.
In peripheral regions, the visual stimulus (image) can be altered without being discriminable from the reference stimulus if the statistics of the image features are close, as quantified by the HVSQ metric (\Eqn{eq:hvsq}).
SP: spatial pooling.
}
    \label{fig:fr_example}
\end{figure}

\paragraph{Foveated Rendering.}
It is well-known that human visual acuity drops as eccentricity increases, i.e., when objects are placed more toward the visual periphery~\cite{wandell1995foundations}.
This is due to a combination of larger pooling sizes~\cite{rodieck1985parasol, dacey1993mosaic} and a sparser photoreceptor distribution~\cite{curcio1990human, Song:2011:ConeDensity} on the retina as the eccentricity increases.
Foveated Rendering (FR)~\cite{patney2016towards, guenter2012foveated} leverages this natural fall-off in visual acuity to speed up rendering by relaxing the rendering quality in peripheral regions. 
\Fig{fig:fr_example} illustrates an example where the visual content in high-eccentricity regions could be altered without being noticeable by users.

While in classic FR the peripherial rendering quality is relaxed by lowering the resolution,
neural rendering offers another dimension: reducing the computational \textit{workload} of each pixel. 
This new dimension is possible because the rendering load of each pixel is controlled by inferencing a deep learning model, which offers many knobs for accuracy-vs-speed trade-offs that have been extensively studied~\cite{han2015learning, blalock2020state, jin2020adabits, yang2019quantization}.
For instance, one can train a smaller model for rendering the visual periphery~\cite{deng2022fov}.
This paper will explore FR knobs unique to PBNR.

\paragraph{Modeling HVS.}
A key question in FR is to determine how much quality to relax without introducing visual artifacts.
It is well-established that commonly used visual quality metrics such as Peak Signal to Noise Ratio (PSNR) or Structural Similarity Index Measure (SSIM)~\cite{hore2010image} do not account for the eccentricity-dependent visual acuity drop in HVS~\cite{walton2021beyond, rosenholtz2016capabilities, strasburger2011peripheral} and, thus, are inadequate for FR: an image with a low PSNR at the visual periphery might not introduce visual artifacts.
The altered image in \Fig{fig:fr_example}, when placed in the visual periphery, is visually indiscriminable from the reference image.

This paper leverages an eccentricity-aware HVS Quality (HVSQ) metric~\cite{walton2021beyond} inspired by classic neuroscience studies about the human visual pathway~\cite{FreemanSimoncelli2011}.
Given a reference image, an altered image, and the eccentricity of each pixel (which depends on the display resolution and the eye-display distance), HVSQ quantifies how similar the two images are as viewed by humans; a lower HVSQ means more similar.

The principle behind the HVSQ metric is as follows.
The retina aggregates photoreceptor outputs in spatial regions, called spatial poolings.
In the image space, a spatial pooling corresponds to a set of adjacent pixels (e.g., SP in \Fig{fig:fr_example}).
The pooling size increases with eccentricity, usually quadratically.
Computational models on HVS~\cite{walton2021beyond} show that as long as the statistics (mean and standard deviation) of the content in a spatial pooling between two images are close, humans can not discriminate between them.
The statistics are calculated in a feature space (as opposed to the pixel space) to emulate the feature extraction in human's early visual processing.

Computationally, the HVSQ of an altered image with respect to a reference image is calculated as follows:
\begin{equation}
  HVSQ = \frac{1}{N}\sum_{i=1}^{N}\Big[\big(\cM(\text{I}^{a}_i) - \cM(\text{I}^{r}_i)\big)^2 + \big(\sigma(\text{I}^{a}_i) - \sigma(\text{I}^{r}_i)\big)^2\Big]
  \label{eq:hvsq}
\end{equation}

\noindent where $N$ is the number of pixels in an image (each pixel has a unique spatial pooling), $\text{I}^{r}_i$ and $\text{I}^{a}_i$ denote the features of the $i^{th}$ spatial pooling in the reference and the altered image, respectively; $\cM$ denotes arithmetic mean, and $\sigma$ denotes standard deviation.

Intuitively, the HVSQ metric calculates the average distance between the two images' statistics across all the spatial poolings.
HVSQ makes intuitive sense: as pixel eccentricities increase, the pooling sizes increase, which gives us more ``wiggle room'' within a spatial pooling to manipulate pixel values to match the feature statistics of the reference image.

\section{Efficiency-Aware Pruning}
\label{sec:prune}

This section introduces a pruning framework to speed up PBNR.
We first identify the root-cause why existing pruning methods are ineffective~(\Sect{sec:prune:perf}).
We then propose two techniques to address the root-cause: intersection-aware pruning (\Sect{sec:prune:metric}) and scale decay~(\Sect{sec:prune:scale}).
Finally, we discuss how these two techniques are combined together~(\Sect{sec:prune:train}).

\begin{figure*}[t]
\centering
\begin{minipage}[t]{0.65\columnwidth}
  \centering
  \includegraphics[width=\columnwidth]{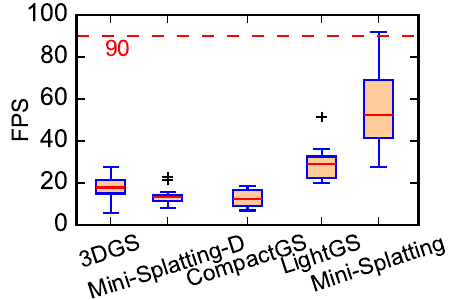}
  \caption{FPS distribution of recent PBNR models on common datasets measured on mobile Volta GPU on Jetson Xavier.}
  \label{fig:fps_boxplot}
\end{minipage}
\hspace{2pt}
\begin{minipage}[t]{0.65\columnwidth}
  \centering
  \includegraphics[width=\columnwidth]{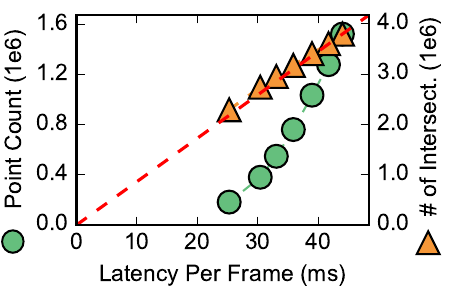}
  \caption{Point count vs. latency per frame and the number of tile-ellipse intersections vs. latency per frame.}
  \label{fig:model_vs_exec}
\end{minipage}
\hspace{2pt}
\begin{minipage}[t]{0.65\columnwidth}
  \centering
  \includegraphics[width=\columnwidth]{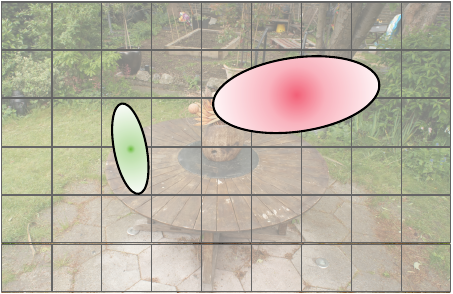}
  \caption{Two ellipses intersect different number of tiles so contribute to computation cost differently.}
  \label{fig:large_small}
\end{minipage}
\end{figure*}

\subsection{Motivations}
\label{sec:prune:perf}

\paragraph{Speed.}
The performance of recent PBNR models is far from real-time on mobile GPUs.
\Fig{fig:fps_boxplot} shows the FPS on the Mip-NeRF 360~\cite{barron2022mip}, Tanks\&Temples~\cite{Knapitsch2017}, and Deep Belending~\cite{hedman2018deep} dataset, measured on the mobile Volta GPU on Jetson Xavier across five recent PBNR models~\cite{fan2023lightgaussian, lee2024compact, fang2024mini, Kerbl2023GaussianSplatting}.
The data is plotted as a standard boxplot to show the FPS distribution across the 13 traces within the datasets.

3DGS~\cite{Kerbl2023GaussianSplatting} and Mini-Splatting-D~\cite{fang2024mini} are two dense models and generally are the slowest.
Much of recent work focuses on pruning: reducing the number of points in a PBNR model~\cite{fan2023lightgaussian, lee2024compact, fang2024mini}.
While effective for reducing the model size, these methods do not significantly speed up rendering.
\revise{For instance, CompactGS, LightGS, and Mini-Splatting} in \Fig{fig:fps_boxplot} are all pruned models;
while generally faster than dense models, they are still \revise{below}
real-time, especially for immersive applications such as AR/VR, which would normally require an FPS of 75--90~\cite{vive_pro2, meta_quest_pro, apple_vision_pro}.

\revise{\paragraph{Why is Existing Pruning Insufficient?}}
Existing pruning methods focus on reducing the point count in a model, which is ineffective for improving speed in PBNR.
To quantify this, \Fig{fig:model_vs_exec} shows the inference latency ($x$-axis) vs. point count (left $y$-axis) of LightGS~\cite{fan2023lightgaussian} (which prunes 3DGS~\cite{Kerbl2023GaussianSplatting}) trained on the \texttt{bicycle} trace in the Mip-NeRF 360 dataset at different pruning levels (between 75\% and 97\%).
The latency reduction rate is slower than that of the point reduction rate.

The reason that reducing the point count is ineffective for acceleration is because the computational costs associated with different points vary.
\Fig{fig:large_small} shows the intuition, where there are two ellipses projected onto the image plane.
The smaller ellipse intersects with only two tiles whereas the larger one intersects with eight.
As a result, the larger one is used in calculating more pixel colors and is naturally responsible for more computation.

Therefore, what \textit{does} impact the inference speed is the number of tile-ellipse intersections.
\Fig{fig:model_vs_exec} shows the latency vs. the average number of intersections per tile (right $y$-axis) for each pruned LightGS model;
the latency reduction rate and intersection reduction rate match.

\subsection{Intersection-Aware Pruning}
\label{sec:prune:metric}

The goal of our pruning is to judiciously reduce tile-ellipse intersections without affecting the visual quality.
The key to our pruning is a metric that we call \textit{Computational Efficiency} (CE), which intuitively describes how much contribution a point makes to pixel values per unit cost of compute.
Intuitively, we would like to prioritize pruning points with low CEs, as they consume a lot of computation without making much contribution to pixel values.
For every point $i$ in a dense model, its CE is defined as:
\begin{equation}
\text{CE}_i = \frac{\text{Val}_i}{\text{Comp}_i}
\label{metric}
\end{equation}

$\text{Val}_i$, contribution of a point $i$ to pixel values, is defined as the number of pixels that are ``dominated'' by that point.
A pixel is dominated by a point if and only if that point, among all the points, has the highest numerical contribution to the pixel value during rasterization (\Sect{sec:bck:pbnr}).
The numerical contribution of a point $i$ is quantified by $T_i\alpha_i$ in \Eqn{eq:p}.

$\text{Comp}_i$, the compute cost of a point $i$, which is ignored in all existing pruning methods, is quantified by the number of tiles that intersect and use (the ellipse of) that point, which directly affects the rendering speed as established above.

In actual rendering, a point will be used in different frames based on the camera pose.
Thus, a point's CE is frame-specific; in extreme cases, a point could be outside the camera's viewing frustum and thus makes no contribution to the image.
We empirically find that the final CE of a point is adequately measured by the maximum CE across all poses (as opposed to the average, which is susceptible to dataset bias) in the training set.

With this metric, during pruning we sort all the points by their CEs and remove a certain portion of points with the lowest CEs.
How many points to remove must be done in conjunction with controlling the quality of the pruned model, which we will discuss in \Sect{sec:prune:train}.

\subsection{Scale Decay}
\label{sec:prune:scale}

Orthogonal to pruning points, another way to reduce tile-ellipse intersections is to reduce the ellipse size/scale, which we call ``scale decay.''
In particular, we want to focus on scaling ellipses that are both large and are used by a lot of tiles in rendering.
To guide scale decay, we propose a metric called \textit{Weighted Scale} (WS) that weighs the point sizes with how often they are used in rendering:
\begin{equation}
\text{WS} = \frac{1}{N}\sum\limits_{i=0}^{N-1} \text{S}_i \text{G}_i
\end{equation}

\noindent where $N$ is the number of points, $\text{S}_i$ is the scale of point $i$'s ellipse (the maximum span of the ellipse in any direction).
Without $\text{G}_i$, WS is simply the average scale of all points in a model.
$\text{G}_i$ weighs a point's scale by how often it is used in rendering, and is defined as:
\begin{equation}
\text{G}_i = (\text{U}_i > T) \cdot ( \text{U}_i - T)
\end{equation}

\noindent where $\text{U}_i$ is the number of tiles a point $i$ is used in rendering and $T$ is a threshold;
intuitively, if a point $i$ is used by fewer than $T$ tiles, its scale is insignificant, in which case the $\text{G}_i$ is 0 so $i$ does not participate in calculating the average scale.
That way, the $\text{G}$ term helps suppressing the scale of points that are not only large but are also used often in rendering.

The $\text{WS}$ metric is a general metric characterizing point/ellipse scales in PBNR.
We empirically find that it is particularly effective when integrated into the training process as an additional term to the loss function $\cL$, which ordinarily is concerned only with the rendering quality ($\cL_\text{quality}$):
\begin{equation}
  \cL = \cL_\text{quality} + \gamma \cdot \text{WS}
  \label{eq:sc-loss}
\end{equation}

\noindent where $\gamma$ is a hyper-parameter governing how much scale decay to apply.

\begin{figure}[t]
    \centering
    \includegraphics[width=\columnwidth]{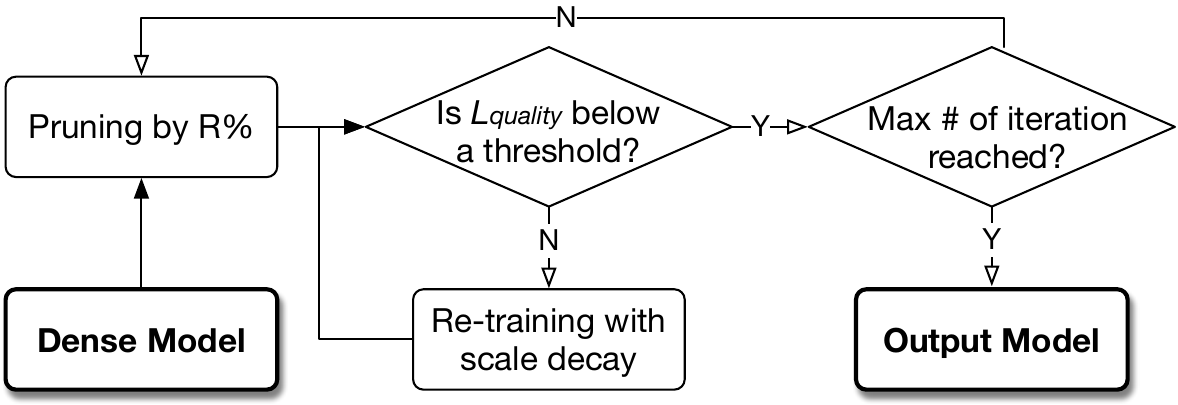}
    \caption{
    The procedure to obtain an efficient PBNR model given a dense model.
    We iteratively apply pruning and re-training with scale decay (guided by $\cL$ in \Eqn{eq:sc-loss}) while controlling for quality ($\cL_{quality}$).}
    \label{fig:mon_loss}
\end{figure}

\subsection{Putting It Together}
\label{sec:prune:train}

Pruning and scale decay are conceptually orthogonal, but, importantly, scaling an ellipse's size also changes its CE.
Thus, scale decay must be done in conjunction with pruning.
\Fig{fig:mon_loss} illustrates the general procedure.

Given a dense model, we first compute the CE for all the points, and repetitively prune a small percentage ($R=10\%$ in our implementation) of the points with the lowest CEs until the quality loss ($\cL_\text{quality}$ in \Eqn{eq:sc-loss}) is above a prescribed threshold.
We then train the pruned model again to regain the quality, but using the composite loss $\cL$ in \Eqn{eq:sc-loss} in order to apply scale decay.
The re-training continues until $\cL_\text{quality}$ is once again below the threshold, at which point we apply intersection-aware pruning again.
We iteratively apply pruning and scale decay in such a way until a certain number of iterations is reached.

Note that $\cL_\text{quality}$ is usually PSNR or SSIM but can be any other quality metric of interest.
In the next section we will show how we can use a human vision-inspired quality metric to account for the eccentricity dependence of visual quality.

Our iterative procedure has two advantages.
First, it combines pruning and scale decay.
Second, it does not require quality-specific hyper-parameter tuning to achieve a specific 
visual quality: monitoring and controlling for $\cL_{quality}$ automatically yield a model at a given quality.

\section{Foveated PBNR}
\label{sec:fr}

\begin{figure*}[t]
    \centering
    \includegraphics[width=2.1\columnwidth]{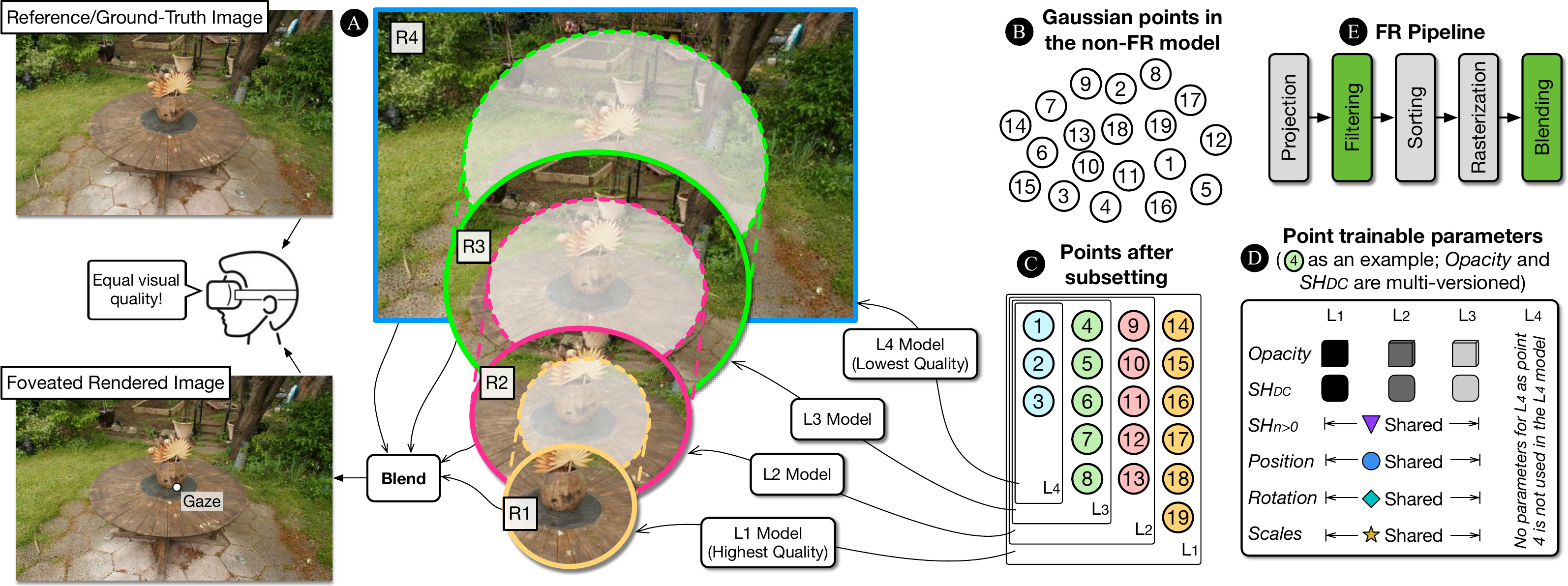}
    \caption{
    The general idea of FR for PBNR.
    \circled{white}{A}: We train multiple models (four in this example), each with a different quality and is responsible for rendering a different quality region in the image (\boxed{R1} -- \boxed{R4}).
    The four quality regions are blended together to generate the final image.
    The goal is for the FR-rendered image to have the same visual quality as the reference image (e.g., generated by a dense model) when judged by humans.
    \circled{white}{B}: Points in the original non-FR model.
    \circled{white}{C}: Our hierarchical point representation to support compute- and data-efficient FR.
    We subset the points so that points used to train a higher-level (lower quality) model are strictly a subset of that used by a lower-level model.
    The \textit{quality bound} $m$ of a point is the highest level that uses the point (e.g., $m=3$ for Point 4).
    \circled{white}{D}: To provide more flexibility for training, we selectively allow key trainable parameters to differ across levels; these parameters are the opacity of a point and the Direct Current (DC) component of the SH coefficients ($\text{SH}_\text{DC}$).
    Other (trainable) parameters of a point are shared across all the levels \textit{that use the point} (e.g., no parameter in $L_4$ for Point 4).
    \circled{white}{E}: The rendering pipeline augmented to support FR (augmentations in green).
    }
    \label{fig:fr_algo}
\end{figure*}

This section introduces a Foveated Rendering (FR) method tailored to PBNR.
We first describe the main idea and its main challenges~(\Sect{sec:fr:mot}).
We then discuss an efficient data representation that enables effective FR for PBNR (\Sect{sec:fr:rep}).
Finally, we describe how to train FR models leveraging the efficiency-aware pruning discussed before (\Sect{sec:fr:train}).

\subsection{Main Idea and Challenges}
\label{sec:fr:mot}

We accelerate rendering by relaxing the rendering quality at the visual periphery, leveraging the low peripheral visual acuity in HVS.
We illustrate the idea in \Fig{fig:fr_algo}, panel \circled{white}{A}.

\paragraph{Main Pipeline.}
As with prior FR work~\cite{deng2022fov, guenter2012foveated, patney2016towards}, we divide an image into $N$ regions (4 in the example), each corresponding to a quality level and is rendered by a separate model.
The region currently under the user's gaze has the highest quality (\boxed{R1} here).
Lower-quality regions are rendered using lighter models, which are obtained by applying pruning and scale decay (\Sect{sec:prune}) to a high-quality model.

Panel \circled{white}{E} shows the rendering pipeline augmented to support FR --- with two new stages (green).
First, after projection we must \textit{filter} each model's points that are outside the model's quality region.
Second, after each region is rendered, we must \textit{blend} the results together to avoid aliasing.

Blending is required in all FR algorithms~\cite{guenter2012foveated, patney2016towards}.
Due to the quality difference across levels, there is a sharp, undesirable boundary between two adjacent levels in the rendered image (a form of aliasing).
To eliminate the boundary, a common technique is for each model to render slightly beyond its assigned boundary;
thus, pixels at the boundary will be rendered twice and then are interpolated/blended to provide a smooth transition between the two levels.

While this multi-model FR idea is conceptually simple, we must address three challenges.

\paragraph{Challenge 1: Performance Overhead.}
FR can potentially accelerate rendering because it reduces the amount of rasterization work in low-quality regions.
However, it has two sources of performance overhead.

First, all $N$ models must go through the Projection and Filtering stages.
In our profiling, these two stages can take up to 18\% of the rendering time.
Second, blending also adds overhead.
Empirically we find that about 25\% of the pixels are to be blended and, thus, rendered twice.

\paragraph{Challenge 2: Storage Overhead.}
FR could increase the model size due to the need to store multiple models, exacerbating the storage pressure of PBNR models.
For instance, the \texttt{bicycle} scene in the Mip-NeRF 360 dataset~\cite{barron2021mip} takes about 1.4 GB of space when trained with 3DGS~\cite{Kerbl2023GaussianSplatting};
recent pruning methods ~\cite{fan2023lightgaussian} reduce the model size of that scene to about 490 MB, which is still large for mobile devices.

We address the first two challenges using an efficient data representation, as we will discuss in \Sect{sec:fr:rep}.

\paragraph{Challenge 3: Controlling Quality.}
FR must be done in a way that guarantees human visual quality --- how do we decide the amount of relaxation at each level?
We describe a training strategy to guarantee consistent human visual quality across all levels, as described in \Sect{sec:fr:train}.

\subsection{Efficient Data Representation}
\label{sec:fr:rep}

\begin{figure*}[t]
\centering
\includegraphics[width=\textwidth]{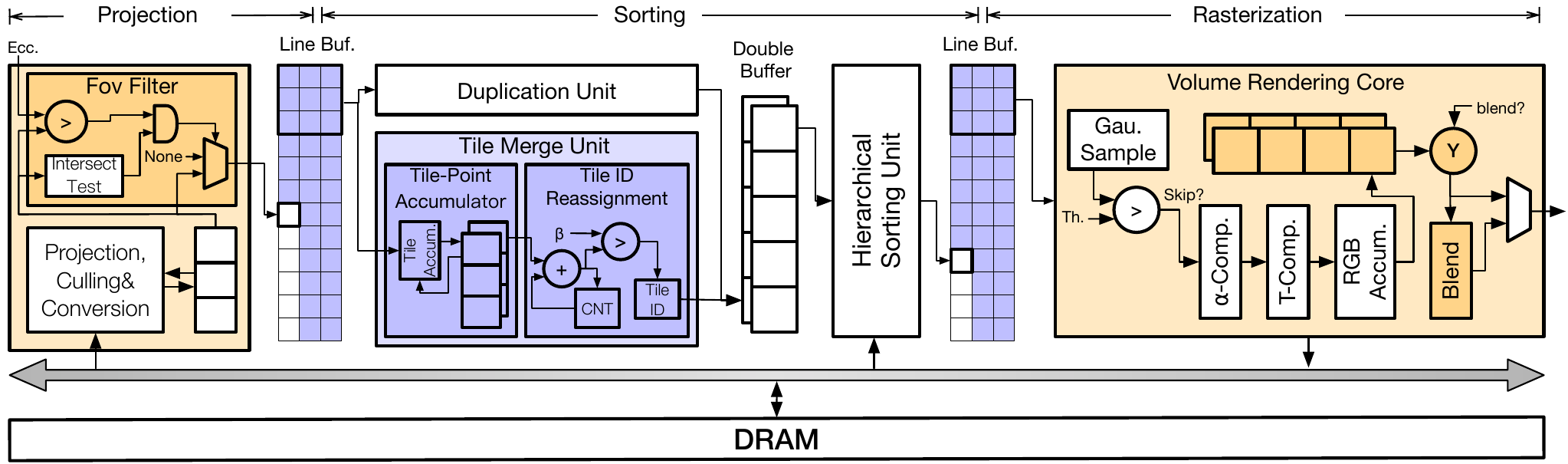}
\vspace{-5pt}
\caption{
The overall architecture design. 
The the basic pipeline is similar to that of GSCore~\cite{lee2024gscore}, a recent PBNR accelerator.
We augment the baseline to support FR (yellow-colored) and to address the workload imbalance issue in PBNR/FR (blue-colored).
}
\label{fig:arch}
\end{figure*}

To address both the performance and storage overhead, we propose an efficient data representation that allows models at different quality levels to share computation and parameters.
The key idea is that points used to train and render a lower quality level are strictly a subset of the points used by a higher quality level.
Panel \circled{white}{C} in \Fig{fig:fr_algo} illustrates how the original points in Panel \circled{white}{B} are organized after subsetting.
Level 1 ($L_1$) model is trained with the most points and thus would offer the highest quality, and Level 4 ($L_4$) model has the fewest points and lowest quality.

Subsetting mitigates both the performance and storage overhead, because the total number of points across all $N$ models, $P_{total}$, is the same as that of the highest-quality model, $P_1$, rather than the sum of all $N$ models.
That is, $P_{total} = \text{max}_{i=1}^N{P_i} = P_1 < \sum_{i=1}^N{P_i}$.
As a result, there is no storage overhead.
The compute overhead is small too, since the Projection and Filtering stages are executed only once, rather than once for each of the $N$ models.

Under subsetting, each point is simultaneously used in models $[L_1, \cdots, L_m]$, where $m$ is the highest level beyond which the point is not used and is called the \textit{quality bound} of the point.
For instance in \Fig{fig:fr_algo}, $m=3$ for Points 4.
During the Projection stage, each point is projected to a tile, which has a specific eccentricity and thus a corresponding quality level $t$.
If $t > m$, the point does not participate in the rest of rendering.
This is the Filtering stage in Panel \circled{white}{E}.

\paragraph{Selective Multi-Versioning.}
Practically, strict subsetting is likely too restrictive in controlling the rendering quality at different levels.
This is because all the trainable parameters of a point would be fixed across all levels, so how a point participates in calculating pixel colors (the $\alpha_i c_i$ term in \Eqn{eq:p}) is also fixed at any time.
In reality, however, a point's contribution to pixel colors should vary depending on the quality region the point is projected to, which varies with the camera pose and the gaze position.

To relax this, we allow multi-versioning as illustrated in panel \circled{white}{D}: a point can maintain $m$ (where $m$ is the quality bound of the point) versions of \textit{some} of its trainable parameters, one version for each level the point is in.
Empirically, we allow four such parameters, i.e., the Opacity and the Direct Current component of the SH coefficients ($\text{SH}_{DC}$); these four parameters are empirically found to impact the pixel colors the most.
We will show in \Sect{sec:eval:fr} that selective multi-versioning is critical to maintain high visual quality.

\subsection{HVS-Guided Training}
\label{sec:fr:train}

The discussion so far has focused on performance, but equally important to FR is the visual quality:
how much weaker can higher-level models be while maintaining subjectively good visual quality across quality levels/eccentricies?

To answer this question, we turn to the HVSQ metric discussed in \Sect{sec:bck:hvs}.
The HVSQ metric quantifies the subjective visual quality between a reference image (e.g., rendered from a dense PBNR model) and an altered image (e.g., rendered from a pruned PBNR model), accounting for the eccentricity-dependent visual acuity of HVS.
Conveniently, while the vanilla HVSQ metric in \Eqn{eq:hvsq} is applied to an entire image, it can be easily adapted to a selected region --- by simply iterating over the spatial poolings (pixels) in the selected region rather than over the entire image.

That way, each quality region has a unique HVSQ measure, and our goal is to ensure the HVSQs across all quality levels are similar to the HVSQ of the baseline model.
To that end, we first train the highest-quality, $L_1$ model, which itself can be pruned and scale-decayed from a dense model.
We then prune a $L_1$ model to obtain a $L_2$ model, which is pruned down to obtain a $L_3$ model; this continues until the desired level is achieved.
The way to obtain a $L_{i+1}$ model follows the exact procedure as laid out in \Sect{sec:prune:train} (i.e., iteratively apply pruning and re-training to a $L_i$ model while controlling for the quality loss $\cL_{quality}$) --- with two key differences.

First, instead of using the usual PSNR/SSIM metrics, we use HVSQ as $\cL_{quality}$ in \Eqn{eq:sc-loss}.
In particular, when obtaining a $L_i$ model we use the HVSQ corresponding to level $i$.
We control for $\cL_{quality}$ so that the HVSQ at all quality levels is the same as that of $L_1$ such that the human visual quality is consistent \textit{across the entire visual field}.
Second, during iterative re-training we do not apply scale decay, because an ellipse scale is not part of the multi-versioned parameters.

\section{Hardware Support}
\label{sec:hw}

Complementing pruning and FR techniques, we propose an accelerator to further improve the performance.
We first provide an overview (\Sect{sec:hw:ov}) and discuss how the hardware addresses the low hardware utilization issue in PBNR, which is exacerbated by FR (\Sect{sec:hw:load}).

\subsection{Overview}
\label{sec:hw:ov}

Our architecture is built on top of GSCore~\cite{lee2024gscore}, a recent PBNR accelerator without FR.
The basic architecture is designed to support the three PBNR stages discussed in \Sect{sec:bck:pbnr}.
The three stages are pipelined across different tiles in a frame.
\Fig{fig:arch} shows the pipelined architecture, with the colored components denoting our augmentations.
The top panel in \Fig{fig:merged_pipeline} illustrates the pipelining process (omitting the Projection stage for simplicity).

\paragraph{Supporting FR.}
We propose two sets of hardware augmentations to support the two new stages in FR (green in panel \circled{white}{E}).
The augmentations are colored in yellow in \Fig{fig:arch}.
First, during the Projection stage we must filter out points that are not used to render a particular quality level.
To that end, we augment the Projection hardware to include a filtering unit, which compares the quality bound $m$ of a tile with the current quality level $t$ of that tile, and pushes the tile to the output buffer only if $t > m$.

Second, FR requires blending across quality levels (\Sect{sec:fr:mot}).
We augment the Rasterization stage with a blending unit, which takes the two colors of a pixel (rendered by two models corresponding to the quality levels) and performs an interpolation.
A small buffer is also needed to temporarily store pixels before blending.

\begin{figure}[t]
\centering
\subfloat[Heatmap showing the number of intersections per tile in \texttt{bicycle}.]
{
    \label{fig:imbalance_exp}
    \includegraphics[width=.47\columnwidth]{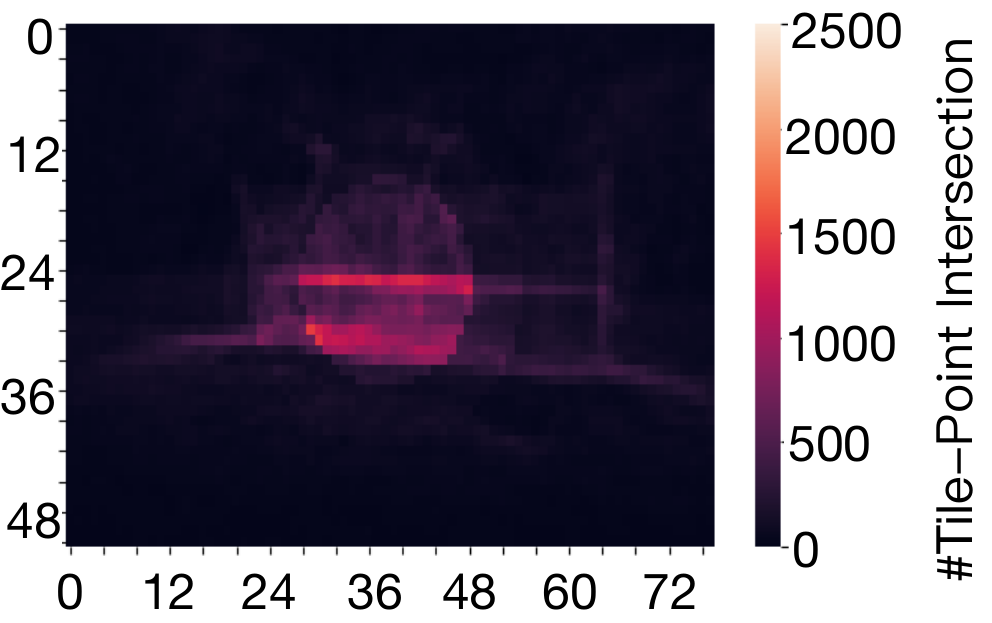}
}
\hspace{2pt}
\subfloat[Boxplot of the intersection distribution in five traces (clipped at 1,500).]
{
    \label{fig:imbalance_across_scene}
    \includegraphics[width=.47\columnwidth]{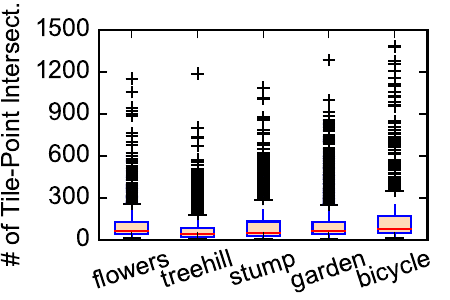}
}
\caption{Workload imbalance, quantified by the number of intersections per tile, on the Mip-NeRF 360 dataset~\cite{barron2022mip}.
The top and bottom notches in the boxplot represent data points that are 1.5 Interquartile Range (IQR) above the third quartile and below the first quartile, respectively.
}
\label{fig:acc}
\end{figure}

\subsection{Addressing Load Imbalance}
\label{sec:hw:load}

\begin{figure}[t]
    \centering
    \includegraphics[width=\columnwidth]{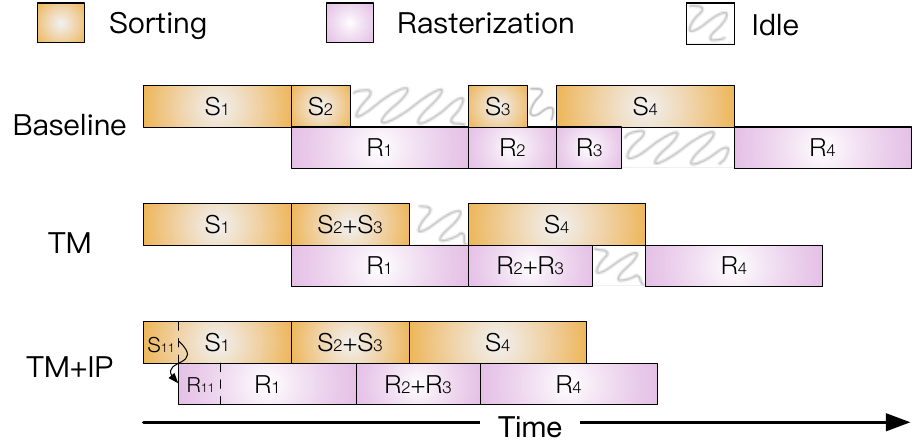}
    \caption{The baseline pipeline faces frequent stalls when the workload is imbalanced across tiles.
    Tile Merging (TM) and Incremental Pipelining (IP) mitigate the workload imbalanced issue and improves the pipelining efficiency.
    With IP, when the first sub-tile $S_{11}$ in the $S_1$ tile is available by the Sorting stage it can be processed by the Rasterization stage.}
    \label{fig:merged_pipeline}
\end{figure}

\paragraph{The Issue.}
With the augmentations above, the hardware can functionally support FR, but faces low hardware utilization.
This is because the amount of work each tile requires is severely imbalanced (recall a tile, like an instruction in a conventional processor, is a basic unit for pipelining; \Fig{fig:merged_pipeline}).

The amount of work a tile involves can be quantified by the number of tile-ellipse intersections.
To quantify this imbalance, 
\Fig{fig:imbalance_exp} is a heatmap plotting the number of intersections each tile (16$\times$16 pixels) has when rendering an image in the Mip-NeRF 360 dataset using our FR model (with four quality levels).
The amount of intersections can vary by over three orders of magnitude.
In particular, most of the intersections are concentrated at the center, because the peripheral tiles are rendered using pruned models, which have fewer points.
\Fig{fig:imbalance_across_scene} is the boxplot showing large the distribution of intersections across all traces in the Mip-NeRF 360 dataset~\cite{barron2022mip};
the imbalance issue is universal.

Load imbalance leads to frequent pipeline stalls.
This is illustrated by the top panel in \Fig{fig:merged_pipeline}, which shows the pipeline dynamics when pipelining Sorting and Rasterization across four imbalanced tiles.
We propose two techniques to address this issue: tile merging and incremental pipelining.
The corresponding hardware components are blue-colored in \Fig{fig:arch}.

\paragraph{Tile Merging.} 
One way to balance the workload across tiles is to merge tiles that have few intersections.
This is dealt with by the Tile Merge Unit (TMU) in the Sorting stage.
Conceptually, the TMU merges incoming tiles into a single tile if the cumulative intersections is below a threshold $\beta$.
The second panel in \Fig{fig:merged_pipeline} shows an example where the second and third tiles are merged, which reduces pipeline stalls and improves performance.

The TMU has two stages.
The first stage processes a Gaussian point by incrementing its counter associated with a specific tile ID, storing the result in a temporal buffer.
When the accumulation for one tile is complete, its total count is forwarded to the second stage, which continuously aggregates incoming tiles, accumulates the intersection counts, and compares the cumulative count against $\beta$.
If this threshold is exceeded, a merged-tile is formed, where each constituting tile is augmented with a merged-tile ID, which is sent along with the native tile ID to the sorting unit.

\paragraph{Incremental Pipelining.}
While Tile Merging reduces the workload imbalances, it does not completely eliminate pipeline stalls, because it is unlikely all the merged tiles have exactly the same number of intersections.

To further enhance pipelining efficiency, we propose to incrementally pipeline data between adjacent stages.
In our baseline PBNR accelerator, a double buffer is placed between two adjacent stages; the consumer stage does not start until the producer stage has finished processing the entire tile.
Our idea is to break the workload of a tile into smaller sub-tiles so that the consumer stage can start working on available sub-tiles before the entire tile is ready from the producer stage.
This works because the workload of each pixel is independent.
This is akin to classic superpipelining in processor design~\cite{mirapuri1992mips, shen2013modern}.
The last panel in \Fig{fig:merged_pipeline} illustrates the benefit of incremental pipelining.

To support such an incremental computation, we replace the double buffers between stages with line buffers (LB)~\cite{hegarty2014darkroom, redgrave2018pixel, hennessy2019pvc, ujjainkar2023imagen}, which, under the surface, is a set of small SRAMs, each of which buffers one row of pixels.
Assume a 16$\times$16 tile size, once 16 rows are produced in the LB, the consumer can starting using them.
The line buffer can be small, since it has to buffer only sub-tiles rather than entire tile.

\section{Experimental Setup}
\label{sec:exp}

\paragraph{FR Training Procedure.}
We use four quality regions whose eccentricity starts at 0\textdegree, 18\textdegree, 27\textdegree, and 33\textdegree, respectively, corresponding to about
13\%, 17\%, 21\%, 49\% of image pixels in these four regions, respectively.

We use \mode{Mini-Splatting-D}~\cite{fang2024mini}, the current-best in quality, as the baseline dense PBNR model.
The $L_1$ model is obtained from the dense model through pruning and scale decay as described in \Sect{sec:prune:train}, with an iteration budget of 50,000, followed by another 5,000 iterations of fine-tuning with HVSQ loss.
The three lower-quality models are obtained from their immediately higher-quality model as described in \Sect{sec:fr:train}, with a 7,500 iteration budget. \revise{Our training time is roughly three times as much as that of the original, dense model.
Much of the slow down is because the HVS loss is calculated using an open-source Python implementation~\cite{hvsq}, which could be accelerated with a more efficient implementation in, e.g., CUDA.}

\paragraph{Variants.}
We design three variants of our method, namely \mode{\proj-H}, \mode{\proj-M}, and \mode{\proj-L}, with decreasing rendering quality.
The $L_1$ model in the three variants is pruned to have a PSNR of 99\%, 98\%, and 97\% of that of the dense model.
The total model size of the three variants is 16\%, 12\%, and 10\%, respectively, of that of the dense model.

\paragraph{Datasets.}
We evaluate three real-world datasets: Mip-Nerf360~\cite{barron2022mip}, Tanks \& Temple~\cite{Knapitsch2017}, and DeepBlending~\cite{hedman2018deep}, which amounts to 13 traces in total.
The camera poses in the datasets are usually very sparsely populated, which is not representative of the continuous rendering scenario (e.g., VR).
We interpolate between the poses in the dataset to create smooth trajectories, producing approximately 1,440 poses for each trace, corresponding to a 16-second video at 90 FPS.

\paragraph{Hardware Implementation.}
We develop a RTL implementation of the accelerator, where the basic pipeline (the uncolored in \Fig{fig:arch}) is similar to GScore~\cite{lee2024gscore},
with the resource allocation adjusted for a more balanced pipeline for our workloads (8 Culling and Conversion Units, a single Hierarchical Sorting Unit, and a 16$\times$16 Volume Rendering Core array).
Our RTL design is implemented via Synposys synthesis and Cadence layout tools in TSMC 16nm FinFET technology.
Each line buffer has a capacity of 1 KB, and the double buffer before the sorting unit is 64 KB.
The SRAMs are generated by an Arm compiler.
The DRAM is modeled after four channels of Micron 16 Gb LPDDR3-1600 memory~\cite{micronlpddr3}.

Overall, we have an area of 2.73 mm$^2$.
The volume Rendering Core takes 63\% of the total area, other stages occupy the rest 30 \%; the SRAMs comprise 7\% of the total area.
Our area is larger than that of GScore (1.45 mm$^2$), whose area is scaled to 16nm using the DeepScaleTool~\cite{sarangi2021deepscaletool}. 
We will show in \Sect{sec:eval:gscore} that we outperform GScore even under the same area when the latter is scaled up.

\begin{figure*}[t]
\centering
\begin{minipage}[t]{0.53\textwidth}
  \centering
  \includegraphics[width=\columnwidth]{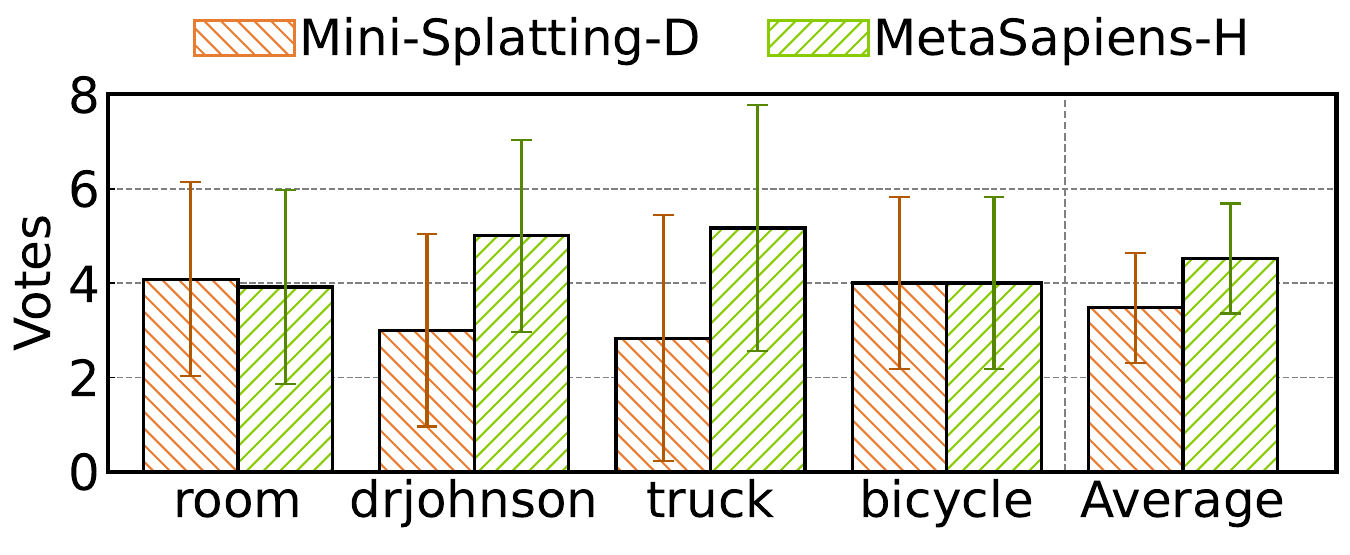}
  \caption{The average number of times the two methods are preferred by users (a tie would be 4-vs-4).
    Error bars indicate the standard deviation within the participants.
    Users either have no preference or prefer our method (binomial test on the average result; $p$ < 0.01).
}
  \label{fig:sub_exp}
\end{minipage}
\hspace{4pt}
\begin{minipage}[t]{0.42\textwidth}
  \centering
  \includegraphics[width=\columnwidth]{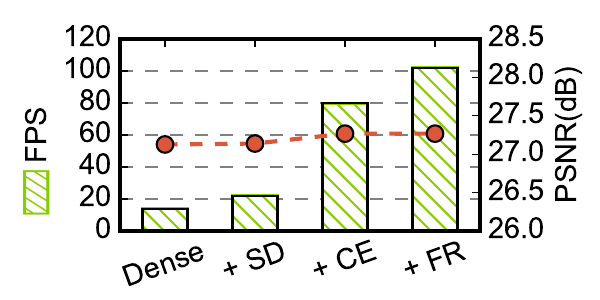}
  \caption{Ablation study teasing apart the impact of various techniques.
  The FPS results are obtained on Jetson Xavier and averaged over all traces.}
  \label{fig:fr_comp}
\end{minipage}
\end{figure*}
\paragraph{Baselines.}
We compare against five recent PBNR models:
\begin{itemize}
    \item Dense PBNR models: \mode{3DGS}~\cite{Kerbl2023GaussianSplatting}, which is the earliest PBNR model, \revise{ and \mode{Mini-Splatting-D}~\cite{fang2024mini}, \mode{Mip-Spla-}\mode{tting}~\cite{yu2024mip}, \mode{StopThePop}~\cite{radl2024stopthepop}, which are state-of-the-art work that improve upon \mode{3DGS}.} %
    \item Pruned PBNR models: \mode{LightGS}~\cite{fan2023lightgaussian}, \mode{CompactGS}~\cite{lee2024compact}, and \mode{Mini-Splatting}~\cite{fang2024mini}. The first two are pruned from \mode{3DGS} and the last one is from \mode{Mini-Splatting-D}.
\end{itemize}

We also compare with two FR methods applied to PBNR.
Both methods use the same quality-region division as in our method.
The first one is \mode{SMFR} (Single-Model FR), which uses a single dense PBNR model, which is the $L_1$ model in \mode{\proj-H}, and randomly samples the points when rendering lower-quality regions.
It is effectively a strict subsetting version of our model without selective multi-versioning.
The second one is \mode{MMFR} (Multi-Model FR)~\cite{deng2022fov}, whose $L_1$ model is the same as that of \mode{\proj-H} and whose higher-level models are pruned from its $L_1$ model separately (without subsetting).
The number of points used in each level in both methods matches that used in our method.

\paragraph{User Study Procedure.}
To assess the subjective rendering quality of our method, we perform a user study;
the procedure is approved by our Internal Review Board (IRB).
We recruited 12 participants (8 males and 4 females between 20 and 30 years old), all with normal or corrected-to-normal vision. \revise{The scale of our user study is comparable with other research in the field~\cite{deng2022fov, rolff2023vrs, shi2024scene, ye2022rectangular}.}
We select four scenes from the three datasets: \texttt{bicycle}, \texttt{room}, \texttt{drjohnson}, and \texttt{truck}, which vary in both content and complexity.
We then render the scenes using our \mode{\proj-H} and \mode{Mini-Splatting-D}, which, recall, is the current-best in rendering quality.

Since \mode{Mini-Splatting-D} does not render in real-time on a mobile device,
we use a workstation with an RTX 4090 GPU to execute both models, both of which render smoothly at 90 FPS.
The workstation streams, in real-time, the rendering to a Meta Quest Pro headset, which has an eye tracker to track the user's real-time gaze.

We use the classic Two-Interval Forced Choice (2IFC) psychophysical procedure~\cite{perez2019pairwise, chen2024pea}, which is commonly used for evaluating the subjective quality of foveated rendering\revise{~\cite{guenter2012foveated, walton2021beyond, deng2022fov, rolff2023vrs, shi2024scene, zhang2024retinotopic, ye2022rectangular, wang2024foveated} }
For each trace, we display its rendering by the two methods on the headset in a random order to each participant, with a 5-second rest interval in-between.
Each participant is then asked to pick which of the two versions they prefer.
Each trace is repeated eight times, and the repetitions across traces are also randomized.
The entire experiment lasts about one hour for each participant.

\section{Evaluation}
\label{sec:eval}

We first show that the subjective rendering quality of our method is statistically no-worse than \mode{Mini-Splatting-D} (\Sect{sec:eval:sub}).
We then show that we provide a better speed-quality trade-off than virtually all baselines on a mobile GPU (\Sect{sec:eval:render}); hardware acceleration improves the speed even further (\Sect{sec:eval:hw}).
We out-perform other FR methods (\Sect{sec:eval:fr}) and a prior PBNR accelerator (\Sect{sec:eval:gscore}).

\begin{figure*}[t]
\centering
\begin{minipage}[t]{0.98\textwidth}
  \centering
  \includegraphics[width=\columnwidth]{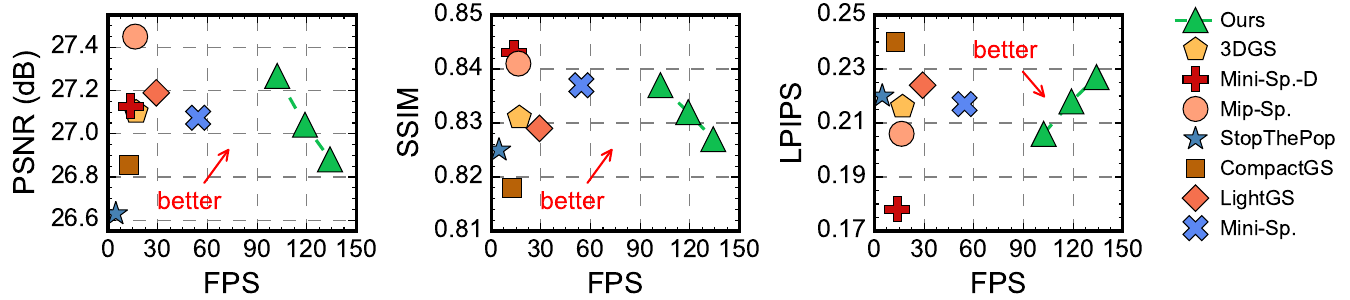}
  \caption{Performance and objective rendering quality (PSNR, SSIM, LPIPS) comparison across the seven baselines and the three \proj variants on the mobile Volta GPU.
  \revise{ \mode{3DGS}, \mode{Mini-Splatting-D}, \mode{Mip-Splatting}, and \mode{StopThePop} are dense models, and the other three baselines are pruned models.}
  }
  \label{fig:gpu}
\end{minipage}
\end{figure*}

\subsection{Subjective Experiments.} 
\label{sec:eval:sub}

\Fig{fig:sub_exp} shows the average number of participants who prefer the two methods for each video.
A tie would be 4-vs-4, since each video is watched eight times by each user.
We find that users either have no preference or prefer our method over \mode{Mini-Splatting-D}.
The results are statistically significant through a binomial test with $p$ < 0.01; the null hypothesis is ``users prefer \mode{Mini-Splatting-D} more than 50\% of the time''.

It might initially look surprising that we have equal or better subjective quality than \mode{Mini-Splatting-D}, a dense model from which we prune and build our FR model.
Further inspection and interviewing participants show two reasons.
First, our HVS-aware fine-tuning (\Sect{sec:exp}) better aligns the statistics of human-sensitive features with the ground truth.
Second, some points in the dense model are trained with inconsistent information across camera poses, leading to incorrect luminance changes over time; pruning those points helps alleviate this inconsistency.

\subsection{GPU Results}
\label{sec:eval:render}

We now show the performance results on the mobile Volta GPU on Nvidia Jetson AGX Xavier~\cite{xaviersoc}, a representative mobile device for use-cases such as VR.
\Fig{fig:gpu} shows the results.
We execute each model five times for each camera pose in each scene in all the datasets, and report the average FPS.

\revise{To put the performance results in context, we compare our variants with the baselines using three objective metrics, PSNR, SSIM, and LPIPS,
which provide good quality measures for the region under the user's gaze and are commonly reported in prior work. }
Using the objective metrics also allows us to scale up the study to more traces.

Our three variants \revise{provide better speed-quality trade-offs in virtually all metrics.} %
Our slowest variant \mode{\proj-H} is 1.9$\times$ faster than the fastest baseline while having better or similar objective quality.
The fastest variant \mode{\proj-L} is 7.9$\times$ faster than \mode{3DGS}, and can be up to 19.8$\times$ on the largest \texttt{bicycle} trace.

\paragraph{Ablation Studies.}
We now ablate the contribution of various performance-enhancing techniques.
\Fig{fig:fr_comp} shows the FPS (left $y$-axis) and PSNR (right $y$-axis) under: 1) the dense \mode{Mini-Splatting-D} model, 2) \proj with only scale decay (SD; \Sect{sec:prune:scale}), 3) \proj with SD and CE-based pruning (\Sect{sec:prune:metric}), and 4) \proj with SD, CE, and FR (\Sect{sec:fr}).
We use the \mode{\proj-H} model and obtain the FPS/PSNR results on Xavier averaged over all traces. 

The PSNRs for all the variants are similar.
With a similar quality, our SD implementation achieves 1.6$\times$ speedup compared to original dense model;
CE-based pruning and FR bring the speedup to 5.8$\times$ and 7.4$\times$, respectively.
CE reduces the model size by 85\%, and FR diminishes the pruning rate only marginally to 84\% owing to selective multi-versioning,.

\begin{figure}[t]
\centering
\begin{minipage}[t]{0.48\columnwidth}
  \centering
  \includegraphics[width=\columnwidth]{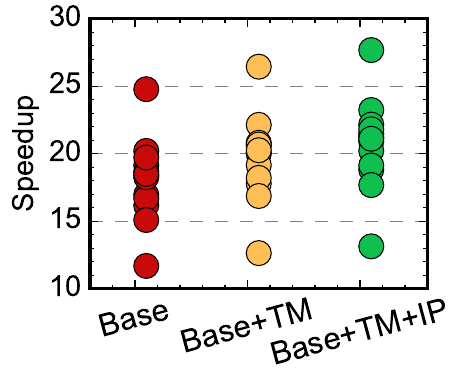}
  \caption{Speed-ups of different accelerator variants over the GPU baseline; each marker is a dataset trace.}
  \label{fig:perf}
\end{minipage}
\hspace{2pt}
\begin{minipage}[t]{0.48\columnwidth}
  \centering
  \includegraphics[width=\columnwidth]{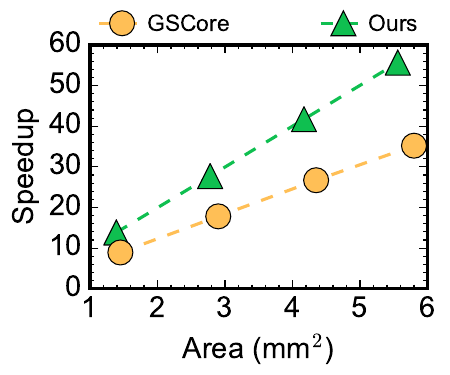}
  \caption{Speedup and area comparison between our hardware and GSCore under different configurations.}
  \label{fig:compare_gscore}
\end{minipage}
\end{figure}
\subsection{Results with Hardware Support}
\label{sec:eval:hw}

\paragraph{Speedup.}
Our hardware support provides further performance improvements.
\Fig{fig:perf} shows the speedups over GPU of: 1) the base accelerator (the uncolored in \Fig{fig:arch}), 2) the accelerator with only Tile Merging, and 3) one with both Tile Merging and Incremental Pipelining.
Each marker represents one of the 13 dataset traces.
We use the \mode{\proj-H} model for evaluation.
Overall, even the base accelerator achieves a 18.5$\times$ speedup (geomean), up to 24.8$\times$, compared to the GPU baseline across different datasets. 

The introduction of Tile Merging consistently improves performance, because tile merging mitigates the load imbalance across tiles. 
On top of that, Incremental Pipelining completely addresses the load imbalance in FR.
Overall, \mode{\proj-TM-IP} combines both techniques and achieves an average 20.9$\times$ (up to 27.7$\times$) speedup.

\paragraph{Energy Savings.}
We summarize the energy results.
Our base accelerator achieves a 54.4$\times$ energy reduction compared to the GPU baseline. 
\mode{\proj-TM-IP} improves the energy saving to 56.8$\times$;
this is primarily because with incremental pipelining we can afford to smaller SRAMs as line buffers, which reduces the energy consumption of SRAMs.

\begin{table} 
\caption{Comparison of FR methods.}
\resizebox{\columnwidth}{!}{
\renewcommand*{\arraystretch}{1}
\renewcommand*{\tabcolsep}{4pt}
\begin{tabular}{ c|cccccc } 
\toprule[0.15em]
 \textbf{Methods} &  FPS $\uparrow$  & Storage (MB) $\downarrow$ &  \multicolumn{4}{c}{ HVS Quality ($\times 10^{-5}$)$\downarrow$ }  \\
 &  & & L1 & L2 & L3 & L4 \\
\midrule[0.05em]
SMFR & 125.9 (1$\times$) & 161.6 (1$\times$) & 2.12& 10.1& 21.7 & 28.3\\
MMFR  & 52.6 (0.42$\times$) & 311.0 (1.92 $\times$)  & 2.12 & 1.87 & 1.79 & 1.76\\
MetaSapiens-H & 102.2 (0.81$\times$) & 171.8 (1.06$\times$)  & 2.12 & 2.10 & 2.09 & 2.08 \\
\bottomrule[0.15em]
\end{tabular}
}
\vspace{-5pt}
\label{tab:fr_comp}
\end{table}

\subsection{Comparison with Other FR Methods}
\label{sec:eval:fr}

\proj also out-performs the two FR baselines.
\Tbl{tab:fr_comp} compares the FPS (on the mobile Volta GPU), storage requirement, and the HVSQ metric across different quality regions/layers.
The results are averaged across all the datasets.

\mode{SMFR} has been seen as a variant of our FR with strict subsetting (no multi-versioning).
Thus, it is the fastest, but has excessively low visual quality, because it simply sub-samples pre-trained points to render low-quality regions.
Its HVSQ in $L_4$ is over 10$\times$ worse than the other two methods.
Users confirm that subjectively this gives the worst quality.

Our method selectively multi-versioned four trainable parameters out of about 60 (\Sect{sec:fr:rep} and \circled{white}{D} in \Fig{fig:fr_algo}), so the additional storage requirement is small (about 6\%).
Note that \mode{\proj-H} already reduces the dense model size to 16\% as shown in \Sect{sec:exp}.

\mode{MMFR} can be seen as a variant of our FR that multi-versions \textit{all} parameters.
It is thus the slowest of the three --- its FPS is way below a 90 FPS real-time requirement, and has the largest storage requirement.
This is due to the compute and storage overhead discussed in \Sect{sec:fr:mot}.
Its HVSQ metrics in higher levels (lower-quality regions) are better than that of ours.
Given that our method is already subjectively no-worse than even a dense model (\Sect{sec:eval:sub}), this suggests that \mode{MMFR} unnecessarily optimizes for details that are imperceptible to users, which we confirm with users.

\subsection{Comparison with GSCore}
\label{sec:eval:gscore}

Our accelerator is based on GSCore~\cite{lee2024gscore}, but has a larger area (\Sect{sec:exp}).
This is because our baseline hardware (uncolored in \Fig{fig:arch}) has 4$\times$ more Volume Rendering Cores compared to that of GSCore with 2$\times$ fewer sorting unit to balance the latency of different stages.

\Fig{fig:compare_gscore} compares the speedup over GPU and area between our architecture (with TM and IP) and GSCore;
both hardware execute the \mode{\proj-H} model on the \texttt{flowers} scene. 
We proportionally scale both GSCore and ours based on their own resource ratio.
We consistently achieve higher speedups with a slightly smaller area against GSCore. 
For example, at an area of around 6 mm², \proj out-performs GSCore by 1.6$\times$.
The higher performance of \proj is from the effectiveness of TM and IP that avoid stalling when pipelining over tiles.
Our performance gain is more significant as the area increases, because the workload imbalance is more pronounced with a large amount of idle resource.

\section{Related Work}
\label{sec:related}

\paragraph{Foveated Rendering.}
The graphics community has long exploited FR for real-time rendering~\cite{patney2016towards, guenter2012foveated, Chen2022InstantReality, Konrad2020GazeContingent, Kaplanyan2019Deepfovea, Krajancich2020Optimizing, Chakravarthula2021GazeContingent, Sun2017PerceptuallyGuided, singh2023power}.
Particularly relevant to our work, researchers have started applying FR to neural rendering~\cite{deng2022fov, rolff2023vrs, rolff2023interactive}, such as Fov-NeRF~\cite{deng2022fov}.
Our work differs from them in two key ways.
First, these methods exclusively focus on NeRF, whereas we focus on PBNR, which is shown to be fundamentally more efficient that NeRF.
Second, some~\cite{deng2022fov} use the multi-model approach, similar to our \mode{MMFR} baseline, which we out-perform (\Sect{sec:eval:fr}).
Our FR method uses subsetting with selectively multi-versioning, addressing the performance overhead of evaluating multiple models (\Sect{sec:fr:mot}).

While conventional FR uses heuristics (e.g., blurring) to guide quality relaxation, recently researchers have investigated more principled ways to model human perception for quality relaxation~\cite{FreemanSimoncelli2011, walton2021beyond}.
This work leverages such theoretical work and integrates it into the training framework to demonstrate its practical utility.

\paragraph{Efficient PBNR.}
Almost all existing work optimizing PBNR focuses on pruning, based on the observation that a considerable amount of points can be pruned without impacting the rendering quality.
They usually do so by, e.g., explicitly training a mask to remove points~\cite{lee2024compact} or sorting points by their numerical contribution to pixel colors followed by removing low-contributing points~\cite{fan2023lightgaussian, girish2023eagles, niemeyer2024radsplat, fang2024mini}.
People have also investigated non-pruning methods, such as vector quantization~\cite{fan2023lightgaussian} and distillation~\cite{lee2024compact} techniques, to compress PBNR models.

Our work differs from them in two key ways.
First, we show that point count is not indicative of performance; tile intersections are (\Sect{sec:prune:perf}).
We propose an intersection-aware metric to guide pruning (\Sect{sec:prune:metric}).
Second, we show an orthogonal technique, scale decay, that complements pruning (\Sect{sec:prune:scale}) and can be performed in conjunction with pruning to further achieve improve performance (\Sect{sec:prune:train}).

\paragraph{Neural Rendering Accelerators.}
Significant work has been done on accelerating neural rendering~\cite{lee2023neurex, rao2022icarus, li2023instant, mubarik2023hardware, li2022rt, fu2023gen, feng2024cicero}, but most of them focuses on NeRF, a variant of neural rendering that PBNR aims to out-perform.
GSCore~\cite{lee2024gscore} accelerates 3DGS, a particular PBNR model.

Our work differs from prior PBNR accelerators in two ways.
First, our hardware supports FR with minimal hardware augmentations.
Second, we identify and address the load imbalance issue in PBNR, which is exacerbated by FR.

\section{Conclusions}
\label{sec:conc}

We achieve over an order of magnitude speedup over existing PBNR models with no subjective quality loss through a user study.
The speedup comes from: 1) a pruning techniques that directly optimizes for the compute-cost of PBNR, 2) FR specialized for PBNR, and 3) hardware support addressing the load imbalanced issue in FR-based PBNR.

\begin{acks}
The work is partially supported by NSF
Award \#2225860.
\end{acks}

\bibliographystyle{ACM-Reference-Format}
\bibliography{references, refs-cicero}

\end{document}